\documentclass{article}

\usepackage{arxiv}

\usepackage[utf8]{inputenc} 
\usepackage[T1]{fontenc}    
\usepackage{hyperref}       
\usepackage{url}            
\usepackage{booktabs}       
\usepackage{amsfonts}       
\usepackage{nicefrac}       
\usepackage{microtype}      
\usepackage{lipsum}
\usepackage{graphicx}
\usepackage{amsmath} 
\usepackage{todonotes}
\usepackage{subfig}

\title{UTLDR: an agent-based framework for modeling infectious diseases and public interventions}

\author{
  Giulio Rossetti \\
  KDD Laboratory\\
  ISTI-CNR\\
  Pisa, Italy \\
  \texttt{giulio.rossetti@isti.cnr.it} \\
   \And
 Letizia Milli \\
  Department of Computer Science\\
  University of Pisa\\
  Pisa, Italy \\
  \texttt{milli@di.unipi.it} \\
  \AND
  Salvatore Citraro \\
  Department of Computer Science\\
  University of Pisa\\
  Pisa, Italy \\
  \texttt{salvatore.citraro@phd.unipi.it} \\
  \And
  Virginia Morini \\
  Department of Computer Science\\
  University of Pisa\\
  Pisa, Italy \\
  \texttt{virginiamorini95@gmail.com} 
}

\begin{document}
\maketitle

\begin{abstract}
Nowadays, due to the SARS-CoV-2 pandemic, epidemic modeling is experiencing a constantly growing interest from researchers of heterogeneous fields of study.
Indeed, the vast literature on computational epidemiology offers solid grounds for analytical studies and the definition of novel models aimed at both predictive and prescriptive scenario descriptions.
To ease the access to diffusion modeling, several programming libraries and tools have been proposed during the last decade: however, to the best of our knowledge, none of them is explicitly designed to allow its users to integrate public interventions in their model.
In this work, we introduce UTLDR, a framework that can simulate the effects of several public interventions (and their combinations) on the unfolding of epidemic processes. 
UTLDR enables the design of compartmental models incrementally and to simulate them over complex interaction network topologies.
Moreover, it allows integrating external information on the analyzed population (e.g., age, gender, geographical allocation, and mobility patterns\dots) and to use it to stratify and refine the designed model.
After introducing the framework, we provide a few case studies to underline its flexibility and expressive power.
\end{abstract}


\section{Introduction}
\label{sec:intro}
Starting from the end of 2019, the Severe Acute Respiratory Syndrome Coronavirus 2 (SARS-CoV-2) has produced an outbreak of pulmonary disease, soon become a global pandemic.
Such a global event - that profoundly affected the behaviors of individuals all over the world - abruptly focused the interest of media and researchers on a pillar field of complex systems research: computational epidemiology.
During the last months, researchers from several fields (as well as the broad population) approached the rudiments of epidemic modeling, searching for a better understanding of the continuously evolving situation and trying at the same time to come out with their prediction for the future.
As reported in a recent survey \cite{estrada2020covid}, countless studies have been proposed approaching SARS-CoV-2 modeling from different perspectives.
Although such a renewed interest in epidemic modeling acts as a valuable linchpin for novel valuable research, a usual barrier often prevents newcomers from testing their ideas: the lack of easy to use tools to implement their models.
Indeed, during the last decades, several programming libraries and visual tools have been released to address such an issue: however, with few notable exceptions, most of them only provide a small set of classic models, not quickly allowing the definition of novel ones.
In particular, almost none of the resources available offers integrated facilities to embed public intervention policies within epidemic modeling (in an easy way).
For this reason, we developed a simple framework, namely UTLDR.

Built on top of NDlib \cite{rossetti2018ndlib}, UTLDR provides a modular backbone that allows defining compartmental epidemic models that incorporates several intervention strategies (quarantine, lockdown, testing and tracking, vaccination\dots) as well as refined information on population stratification and human mobility.
Our framework differentiates the meta-compartments in which individuals can transit among five groups: Undetected, namely the non identified exposed/infected individuals; Tested, the infected individuals that are identified and followed by the healthcare system; Lockdown, the individuals that are subject to social distancing and mobility restrictions; Recovered and Dead, those individuals that completed (either with/without partial immunity in case of Recovered ones) their path.
Moreover, UTLDR also provides several extensions to cover additional intervention strategies and sanitary risks (e.g., vaccination campaigns, inefficient corpse disposal\dots).

Although not explicitly designed for SARS-CoV-2 modeling, UTLDR can be easily used to simulate diffusive scenarios and represent a starting point for advanced epidemic modeling.
Conversely, from recent studies, our aim is not to describe a specific model but to allow UTLDR users to define their own. 
UTLDR is designed to both facilitate newcomers in testing their ideas and the general public - without a strong scientific background - to play with a simulator and, hopefully, gain consciousness of both challenges of epidemic modeling and reasons (as well as potential effects) behind standard non-medical interventions.

The paper is organized as follows. 
Section \ref{sec:related} briefly analyzes the relevant literature on compartmental epidemic modeling and reviews the existing tools designed to perform diffusion simulations on top of network structures.
In Section \ref{sec:model} is described, in an incremental fashion, the UTLDR framework by proposing a few examples on how its components can be combined together to include different public interventions strategies. 
In Section \ref{sec:casestudy} a few models built with the UTLDR framework are tested against (i) synthetic social interaction networks, and (ii) interacting agents stratified to match the population of an Italian region, Tuscany.
Finally, Section \ref{sec:future} concludes the paper.

\section{Related Works}
\label{sec:related}
In this section, we report a brief overview of the founding literature on compartmental epidemic modeling and of the libraries and tools available to researchers.
Additional references that are relevant to our contribution will be provided in Section \ref{sec:model} while introducing UTLDR.

\noindent{\bf Compartmental Models.} A great reduction of a system to understand spreading is a compartment model (\cite{pastor2015,newman2002spread,hethcote2000mathematics,heesterbeek2000mathematical,anderson1992infectious}).  
The basic idea of these models is to divide the population into disjoint groups (compartments), according to a few key characteristics which are relevant to the process under consideration, then the  evolution of an epidemic is modeled by keeping track of the number of individuals within each compartment. 
This approach relies on the assumption that populations are fully mixed, meaning that people interact with each other at random and each member in a compartment is treated indistinguishably from the others in that same compartment. 
These interactions, and in general transition processes between the compartments, are captured in the model as in the limit of large population size: thus, we can fully specify them with nonlinear differential equations accounting for the changes in the number of individuals in the various compartments.

A basic compartmental model that is applicable to many common infections is the SIR model, where we divide the population into those who are
susceptible ($S$), those who are infected ($I$) and those who have recovered and are immune ($R$). In the simplest setting, the total population is assumed constant, but extensions with demographic processes are common. The corresponding system of ordinary differential equations reads
\begin{equation*}\aligned
\frac{dS}{dt}&= -\beta S(t) I(t),\\
\frac{dI}{dt}&= \beta S(t) I(t)-\gamma I(t),\\
\frac{dR}{dt}&= \gamma I(t),
\endaligned
\label{eq:SIR}
\end{equation*}
where $S(t)$, $I(t)$, and $R(t)$ are the fractions of the population
in each of the three states at time $t$ (\cite{pastor2015epidemic,newman2002spread}). Under the assumption of a fully mixed population, $\beta$ is the average rate of infective individuals that have contacts with other individuals per unit time, and $\gamma$ is the recovery rate. In the special case when $\gamma=0$, we arrive at the SI model that assumes that individuals never recover from the infection. On the other hand, to model diseases where recovery does not confer immunity, one easily turns the above system into a SIS model, by moving the term $\gamma I(t)$ into the right hand side of the first equation and canceling the third equation. Numerous variants of the SIR model have been devised in the literature, for example by specifying
further compartments such as those who have been vaccinated, those who are receiving treatment, age groups, risk groups, etc. Other models arise as we consider a different term for the transmission process, or assume different distribution for the time individuals spend in the infected compartment, leading to a non-constant recovery rate. Spatial effects
can be incorporated by adding diffusion terms to the equations, or by considering patch models and the underlying network of individuals' mobility. 
\\ \ \\
\noindent{\bf Libraries and Simulators.}
When it comes to model and study complex networks and diffusive phenomena, several resources are available to students, programmers and researchers.
Indeed, our review will not cover all the existing resources but only the ones that, in our opinion, provide interesting facilities to the end user at a reasonable learning cost.

One of the main library designed to handle, manipulate and analyze graph structures in R is \emph{Igraph}\footnote{Igraph: \url{http://igraph.org/redirect.html}}\cite{csardi2006igraph}.
Igraph is written in C and is released as Python and R packages. 
It can handle large graphs very well and provides functions for generating random and regular graphs, graph visualization, centrality analysis, path length and much more.
When it comes to simulating epidemic models in R one of the most famous package is undoubtedly \emph{EpiModel}\footnote{EpiModel: \url{http://www.epimodel.org/}}\cite{epimodel}.
EpiModel provides facilities for build, solve, and plot mathematical models of infectious disease. 
It currently provides functionality for three classes of epidemic models -- Deterministic Compartmental Models, Stochastic Individual Contact Models and Stochastic Network Models -- and  three types of infectious disease can be simulated upon them: SI, SIR, SIS. 
EpiModel allows generating visual summaries for the execution of epidemic models; it provides plotting facilities to  show the means and standard deviations across multiple simulations while varying the initial infection status. 
It also includes a web-based visual application for simulating epidemics\footnote{EpiModel viz: https://statnet.shinyapps.io/epinet/}.

The most famous, pure Python package, that provides graph data structures along with algorithms, synthetic generators and drawing tools is for sure \emph{NetworkX}\footnote{NetworkX: \url{https://networkx.github.io}}\cite{Hagberg2008}. 
Upon such general graph modeling framework is built the \emph{Nepidemix}\footnote{Nepidemix: \url{http://nepidemix.irmacs.sfu.ca/}} library: a suite  tailored to programmatically describe simulation of complex processes on networks \cite{ahrenberg2016nepidemix}. 
\emph{Nepidemix} was developed by members of the IMPACT-HIV group; it is written in Python 2 and uses the module NetworkX to manage the network structure. 
It automates common diffusion simulation steps allowing the programmer to build a network according to some specifics and to run on top of it a set of epidemic processes for a specified number of iterations. 
Moreover, Nepidemix allows during simulation to save incremental results such as disease prevalence and state transitions.
Another Python library dedicated to the simulation of diffusive models is \emph{EoN}\footnote{\url{https://github.com/springer-math/Mathematics-of-Epidemics-on-Networks}}. 
EoN is designed to study the spread of SIS and SIR diseases in networks \cite{211986}.
It is composed of two sets of algorithms: the first set that deals with simulation of epidemics on networks (SIR and SIS) and the second designed to provide solutions of systems of equations. 
Finally, a recent and easily easily extensible library has been proposed in \cite{rossetti2018ndlib}. 
NDlib\footnote{NDlib documentation: \url{ndlib.readthedocs.io}} offers support to a vast ensamble of diffusion models both coming from the Epidemic literature and the Opinion Dynamic one. 
Due to its standardized interface and flexibility we used NDlib as the backbone on top of which implementing UTLDR\footnote{UTLDR implementations details available at \url{https://bit.ly/362xfe7}}.

\begin{figure}[t]
\centering
 \subfloat[]{\includegraphics[scale=0.22]{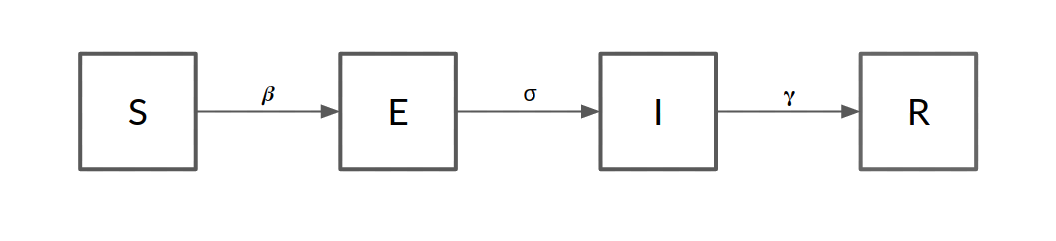}}
  \subfloat[]{\includegraphics[scale=0.40]{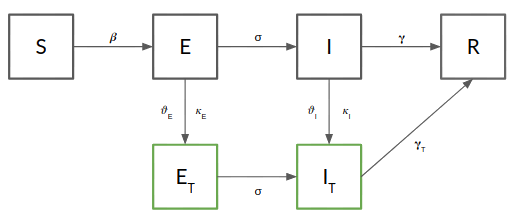}}

  \caption[Diffusion models: SEIR model]{(a) In the SEIR model an individual can be in one of four states: (S)usceptible, (E)xposed, (I)nfected or (R)emoved. Arrows indicate transitions among compartments. (b) The UTR model extends SEIR by introducing the Tested meta-compartment (blocks in green). Testing can be applied to both Exposed and Infected populations and results in transitions to Quarantine compartments ($E_T$ and $I_T$).}
\label{SEIR}
  \end{figure}

\section{UTLDR framework incremental description}
\label{sec:model}

The framework that we propose is built as a conservative extension of the SEIR one \cite{aron1984seasonality}, designed to organize the population in five meta-statuses: (U)ndetected, (T)ested, (L)ockdown, (D)ead, (R)ecovered.

In this section, we will provide, at a high level of abstraction, a description of how alternative/complementary interventions can be (incrementally) added to such a base model to describe the modules of our framework.
Our discussion will focus on the model parameters exposed by each proposed SEIR extension as well as the rationale behind the novel compartments and the transition rules we define.
The proposed extensions can be composed to reach all the possible combinations: in the following, we incrementally discuss them, exemplifying their rationale by using the standard representations offered by transition diagrams.
For a detailed mean-field description of UTLDR, refer to Appendix \ref{appendix}.

\noindent{\bf Base model: SEIR.}
The SEIR model was introduced in 1984 to investigate the role of seasonality in cycles of recurrent epidemics. 
We can suppose that a population can assume four states:  Susceptible ($S$), those individuals able to contract the disease; Exposed ($E$), those who have been infected but are not yet infectious; Infected ($I$), those capable of transmitting the disease; Recovered ($R$), those who have become immune or deceased.

Indeed, many diseases have a latent phase during which the individual is infected but not yet infectious. This delay between the acquisition of infection and the infectious state can be incorporated within the SIR model by adding a latent/exposed population, $E$, and letting infected (but not yet infectious) individuals move from $S$ to $E$ and, only then, from $E$ to $I$.
SEIR assumes that if during a generic iteration, a susceptible individual comes into contact with an infected one, it becomes infected after an exposition period ($1/\sigma$) with probability $\beta$, then it can switch to removed with probability $\gamma$ (the only transition allowed are S$\rightarrow$E$\rightarrow$I$\rightarrow$R). Figure \ref{SEIR}(a) shows the transition diagrams of a classic SEIR model.

\noindent{\bf Testing, Tracing and Quarantine.}
In the absence of specific therapeutic drugs or vaccines for the novel disease, it is essential to detect the diseases early and immediately isolate the infected individual from the healthy population (quarantine).
Quarantine management is a crucial measure that has to be taken once the human-human transmission is confirmed.

So, we generalize the SEIR model by introducing the testing performed on exposed and infected people and the quarantine compartments. We add two statuses (as shown in the transition diagram reported in Figure \ref{SEIR}(b)):
\begin{itemize}
    \item Identified Exposed ($E_T)$: the exposed population that has been identified by testing strategies;
    \item Identified Infected ($I_T$): the infected population that has been identified by testing strategies.
\end{itemize}
We consider the population that reach either of the two statuses (marked in green in Figure \ref{SEIR}(b)) as quarantined.\\
The transitions $E\rightarrow E_T$ and $I\rightarrow I_T$ are regulated by the following parameters:
\begin{itemize}
    \item Testing probability: $\vartheta_E, \vartheta_I$
    \item Testing success rate: $\kappa_E, \kappa_I$ 
\end{itemize}

  \begin{figure}[t]
\centering
 \subfloat[]{\includegraphics[scale=0.35]{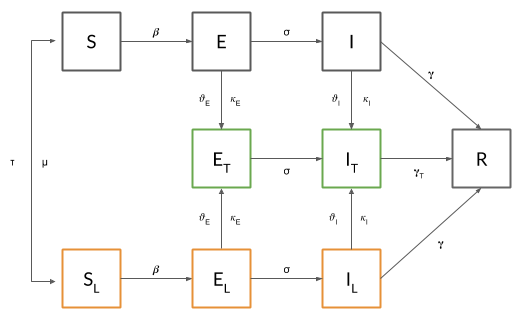}}
 \subfloat[]{ \includegraphics[scale=0.35]{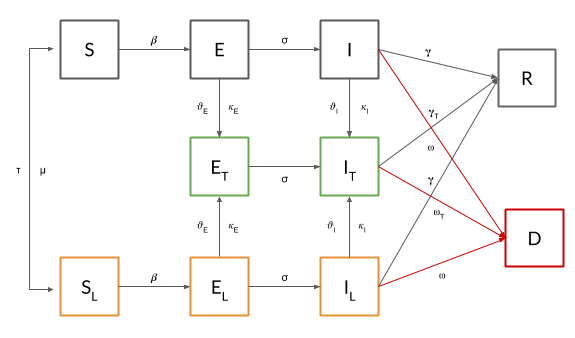}}
  \caption[Diffusion models: UTLR model]{(a) The UTLR extension includes lockdown compartments (blocks in orange). Restrictions can be applied to Susceptible, Exposed and Infected populations and result in transitions to locked compartments ($S_L$, $E_L$ and $I_L$) (b) The UTLDR compartment adds the possibility of differentiating between recovered and immunized population ($R$), and dead ($D$).
  }
\label{UTLR}
  \end{figure}

Therefore, exposed and infected individuals are tested with probability $\vartheta_E, \vartheta_I$, respectively, and each test produces a false positive result with probabilities $\kappa_E, \kappa_I$, respectively. 
Testing positive moves an individual into the appropriate detected case compartment.
Moreover, UTLDR also allows, in case of positive testing result, to enable contact tracing procedures (while specifying a temporal window, $T_{tracing}$, to limit the search)\footnote{Tracing is implemented in the activity driven agent-based version of UTLDR}.

While in a quarantine compartment, individuals are not allowed to infect susceptible ones.
Finally, to capture different recovery rates between $I$ and $I_T$, respectively (assuming quarantined individuals being treated with appropriate medical care), we introduce the $\gamma_T$ as a parameter regulating the $I_T\rightarrow R$ transition.

\noindent{\bf Lockdown and social distancing.}
Another intervention procedure to control the spread of infectious diseases is to reduce individuals' social interactions.
The rationale for social distancing/lockdown strategies is that they slow the spread of the disease (in extreme scenarios, limiting it to individual households), smoothing the infection trend, reducing the pressure on the health care system, and finally, buying time for its strengthening.

To simulate lockdown effects, we add in UTR three statuses (see the transition schema in Figure \ref{UTLR}(a), new statuses marked in orange):
\begin{itemize}
    \item Susceptible in lockdown ($S_L)$: the susceptible population adhering to the lockdown; 
 \item Exposed in lockdown ($E_L)$: the exposed population adhering to the lockdown; 
  \item Infected in lockdown ($I_L)$: the infected population adhering to the lockdown.
 \end{itemize}

Each individual can move from its current status (if in $S$, $E$, or $I$) to the corresponding lockdown status as regulated by the following parameters:
\begin{itemize}
    \item Adherence to lockdown ($\tau$): the probability that an individual adheres to the required lockdown policy;
    \item Lockdown escape probability ($\mu$): inverse of the expected duration of the lockdown; during each iteration, socially distanced individuals can decide to leave the lockdown with a probability $\mu$.
    \end{itemize}

Also, in this case, exposed and infected individuals in lockdown are tested with probability $\vartheta_E, \vartheta_I$, respectively, and a probability of false positive result of $\kappa_E, \kappa_I$, respectively. 
   
To make UTLR more general, we also add to the framework the possibility of considering two different disease outcomes: recovering (and immunization, $R$) or death ($D$).
With such a new compartment, we get the UTLDR model\footnote{Without lack of generality we name the model with all meta-compartments enabled as the general framework. Please, note that the full definition of a model is provided by the values of its parameter, thus there might exists several UTLDR models having a different set of active compartments/transition rules.}, whose transition diagram is shown in Figure \ref{UTLR}(b).

Since the added final compartment can be reached from all the infected ones (namely, $I$, $I_T$, and $I_L$), we extend  UTLDR with different transition probabilities to regulate its incoming transition rules.
In particular, we model with $\omega$ the real lethality rate (that regulates the $I\rightarrow D$ transition) and $\omega_T$ the observed one (that regulates the transitions $I_T\rightarrow D$ and $I_E\rightarrow D$.

\begin{figure}[t]
\centering
\subfloat[]{ \includegraphics[scale=0.31]{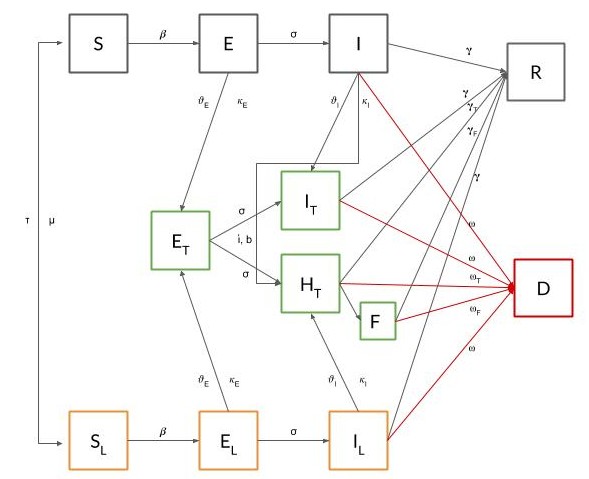}}
 \subfloat[]{\includegraphics[scale=0.31]{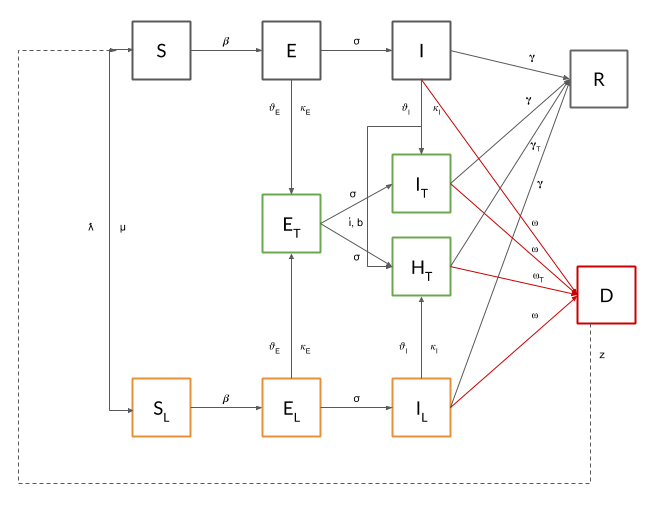}}
  \caption[Diffusion models: UTLDR model]{(a) The UTLDR module with limited ICU availability includes the severe hospitalized population $H_T$, that is differentiated to the one with mild symptoms $I_T$; moreover, population in $H_T$ that can not be treated adequately is placed in $F$. (b) The UTLDR module with corpse disposal allows the contact of population in $S$ with an infected corpse in $D$.}
\label{UTLDRICU}
  \end{figure}

\noindent{\bf ICU with/without availability limits.}

So far, we considered the "tested" compartments (namely $E_T$ and $I_T$) as a proxy to model quarantine.
Indeed, we can extend such characterization to enhance their expressiveness.
In particular, we can maintain the quarantine semantics for the $E_T$ status while leveraging $I_T$ as a first building block of another family of compartments: Hospitalization.

Indeed, the diffusion of some diseases can determine the increasing demand for critical care affected by medical devices' scarcity. 
Such limitations occur, for instance, when there are no available ICU beds for patients with a critical illness, leading to delays in ICU admission that have significant clinical consequences.  Admission delays can result in the boarding of critically ill patients in the emergency area or in other hospital units, which is associated with increased mortality.

To capture such an aspect, we extend our framework with the compartment $H_T$, where tested individuals are hospitalized in case of severe illness - thus making of $I_T$ the compartment collecting ill individuals experiencing mild symptoms (see the extended transition diagram in Figure \ref{UTLDRICU}(a)).

Adding the $H_T$ compartment requires the definition of novel parameters: $\iota$, namely the probability of a severe case (e.g., requiring ICU) that regulates the transition $I_T\rightarrow H_T$.
Moreover, due to lack of medical devices, it could happen that severe cases cannot be placed in $H_T$ and are allocated to standard hospitalization routines, $F$: as expected, the transition $H_T\rightarrow F$ is subject to an upper limit on the resources that we identify as $b$.
The $F$ compartment models severe cases that are not adequately treated and thus potentially affected by a higher lethality ($\omega_T$ instead of $\omega$) and lower recovery probability ($\gamma_T$ instead of $\gamma$).

\noindent{\bf Corpse Disposal Efficiency, Partial Immunity and Vaccination.}
It is well known that in the case of some diseases, such as Ebola \cite{nistal2019new}, the infective lying corpses are infective. The dead infective corpses can be considered in the model as a new sub-population $D$ that can infect again.
In this case, to extend our model, we add the parameter $z$ - the probability of infection from corpses - to regulate the transition $S\rightarrow I$ in case of direct (and leaky) contact of a Susceptible individual with an Infected corpse (Figure \ref{UTLDRICU}(b)).

\begin{figure}
\centering
 \subfloat[]{\includegraphics[scale=0.31]{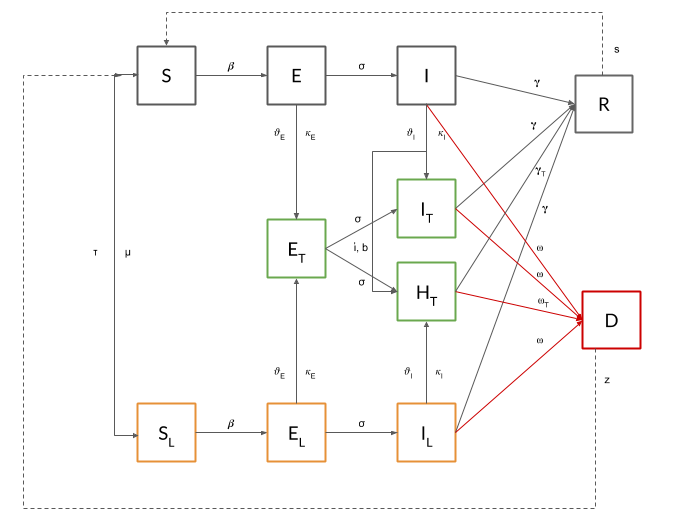}}
 \subfloat[]{\includegraphics[scale=0.31]{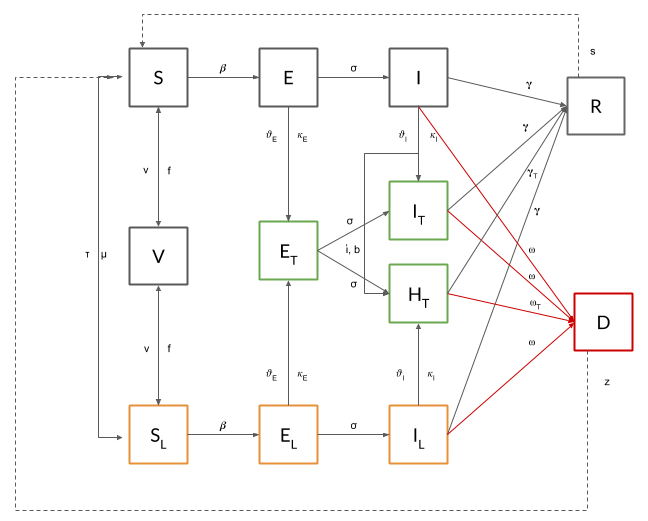}}
  \caption[Diffusion models: UTLDR model]{(a) The UTLDR model with partial immunity considers the possibility that population in $S$ can be reinfected again. (b) The UTLDR model with vaccination includes a 
  new sub-population $V$ that has (successfully) being vaccinated.}
\label{UTLDR2}
  \end{figure}

So far, we have assumed people have lifelong immunity to disease upon recovery.
Indeed, such characteristics cannot be assumed for all possible diseases. 
An individual's immunity might decrease over time, or that a subset of the recovered population can not produce the antibodies for the disease. 
To cover these scenarios, we extend the proposed model allowing recovered individuals to return to a susceptible state, thus allowing the $R\rightarrow S$ transition, under a re-infection probability $s$ - as shown in Figure \ref{UTLDR2}(a).

Finally, the last building block of the proposed framework regards the activation of vaccination campaigns. 
To such extent, we introduce a new compartment, $V$, collecting the sub-population that has (successfully) received the vaccination - as shown in Figure \ref{UTLDR2}(b).
We assume that only susceptible individuals (either in $S$ or in $S_L$) can be vaccinated, imposing a vaccination probability $v$ and a probability of vaccination nullification (inverse of the vaccine's expected temporal coverage) of $f$.
In the case of vaccination nullification, the $V\rightarrow S$ transition is applied.

\section{Extending UTLDR: Agent-based modeling and Human Mobility}
Although designed to capture different stages of infectious disease dynamics, the framework described so far assumes a fully mixed population: every individual in the population is equally likely to interact with every other individual, and each member in a compartment is treated indistinguishably from the others of that same compartment.
Such an approach is widely adopted in epidemic modeling literature; however, it suffers from a relevant limit: it makes model simulations fully deterministic (once fixed the population and model parameters values).

Even though such a simplification allows for a closed analysis and characterization of epidemic models, it is of utmost importance to consider those stochastic effects introduced by the heterogeneous structure of contact networks.
To such extent, we designed the proposed framework to exploit the available information (if any) on the population social tissue, thus transforming individuals into a (possibly) stratified population of \emph{agents} (in the following we use the terms individual/agent interchangeably).

We model agents' social circles as a node-attributed graph $G=(V,E,A)$ where $V=\{V_1,V_2,\dots,V_n\}$ is the set of nodes (agents), $E=\{(u,v)|u,v \in V\}$ is the set of edges (the social ties connecting agents), and $A$ is the set of node attributes (identifying both the UTLDR compartments and node characteristics).
We assume that in the defined feature-rich \cite{interdonato2019feature} interaction graph each agent is fully specified by a set of arguments from $A$, some of them mandatory (e.g., the node's current compartment), other optional (e.g., age, gender\dots). 
For sake of simplicity we allow only the node's compartment attribute to vary during simulation.

Moreover, to better simulate the dynamic nature of social interaction, we assume that not all social ties of a given node are active during each simulation iteration.
To implement such a constraint, we leverage a simple Activity Driven \cite{perra2012activity} network model, a framework often employed to simulate evolutive dynamics of network topology in the absence of explicit temporal interaction data \cite{liu2014controlling,pozzana2017epidemic,zino2018modeling,ogura2019optimal}. 
Each agent $v\in V$ in the network has assigned an activation probability $a_v \in [0,1]$ identifying the percentage of edges (chosen uniformly at random) he activates during each simulation iteration.
Moreover, we also allow part of such interactions to occur outside the neighborhood of $v$. To do so, we augment the model with a probability $p$ (evaluated once for each interaction) to account for long-range contacts.
In particular, we allow each agent to interact with random ones from their neighborhood with probability $(1-p)$ and with random ones from anywhere in the network with probability $p$.
Long-range interactions are introduced to model the chance that agents interact with infected people outside their neighborhood (e.g., while on public transportation or at the supermarket). 
Indeed, the parameter $p$ defines the network's locality: for $p=0$ an agent interacts only with their social circle, while $p=1$ represents a uniformly mixed population. 
The $a_v$ and $p$ parameters are vital factors in simulations involving quarantine/hospitalization and lockdown compartments: the former one is implicitly used to restrict agents' sociality during quarantine/hospitalization, the latter one, to tune account for decreased mobility during lockdowns.

Starting from such a network refined contact model, we defined two alternative versions of the framework, each one assuming a different available knowledge: (i) explicit and (ii) implicit social tissue models.
\\ \ \\
\noindent{\em Explicit Network Structure.}
In this scenario, the social topology connecting individuals is known apriori. 
Explicit social interaction topologies are often obtained from online social network platforms or built on small/medium-scale sensor tracking experiments.
The main issue in working with explicit network structures - apart from being difficult to obtain - lies in the space consumption that grows rapidly as the population size increases.
\\ \ \\
\noindent{\em Implicit Network Structure.}
In this scenario, the social topology is unknown to the analyst: we assume available (or, at least, partially inferable) other information characterizing the population (e.g., workplace, school attended, household size\dots).
Leveraging such external knowledge, as already done in literature (e.g., the model introduced in \cite{ferguson2020report}), we build several social contexts for each agent. From each of them, randomly sample with probability $a_v$ the interactions occurring during each simulation iteration (and with probability $p$ interactions outside them).
This approach efficiently addresses the memory issue in storing the explicit interaction graph (which is now generated on the fly) while introducing higher variability on individuals' contacts.

It is essential to underline that the provided implementation of the UTLDR framework also allows us to stratify all the discussed parameters to characterize the population better if needed.
So far, we suggested that each of the compartments controlling parameters may be specified as (fixed) unique values: we opted for such simplification to ease the incremental framework introduction.
However, when additional knowledge on the studied population is available (e.g., age distribution, household distribution size, gender distribution, employment type/workplace/school size distributions\dots), the proposed framework allows to use it to stratify controlling parameter values directly (e.g., imposing $\beta=0.02$ for female under 18 years old agents while setting $\beta=0.3$ for the rest of the population).
Such flexibility makes it possible to define sub-populations characterized by different responses to epidemic events easily.

Another key component profoundly affecting epidemic spreading is {\em human mobility} \cite{barmak2011dengue,espinoza2020mobility,kraemer2020effect,cintia2020relationship}.
So far, we simplistically modeled long-range interactions with a simple probability; however, such a parameter does not control the actual mobility constraints that might affect random social interactions.
For this reason, we incrementally extended our framework to leverage aggregated mobility information (when available) to select the most likely cohort of agents for short/medium/long-range interactions.

To do so, UTLDR requires three different, additional sources of data: 
\begin{itemize}
    \item aggregated geographic allocation of the population (e.g., as inferable from census data);
    \item  a geographic tesselation  (e.g., a hierarchy composed by census cells, municipalities, regions\dots);
    \item a set of aggregated origin-destination matrices (one for each level of the tesselation hierarchy, e.g., as computable from GPS/CDR data \cite{jain1999estimating,alexander2015origin}) each providing the probabilities of moving from/to any given geographic area to all the others of the same hierarchy level.
\end{itemize} 
In the presence of such knowledge, UTLDR will: (i) allocate each individual to several geographic regions, one for each social circle he/she is involved in (e.g., one for the neighborhood of the home location - as identified by stratification of the population over census cells - and one for the workplace), (ii) sample social interactions among individuals associated to shared social/geographical clusters (weighting them in different ways if needed), and (iii) sample long-range interaction within geographic clusters reached with probability given by the provided origin-destination matrices - assuming as starting location the individual's home one.

\section{Case studies}
\label{sec:casestudy}
In this section, we report a few case studies to underline the proposed UTLDR framework's flexibility.
In particular, without losing generality, we propose two case studies: a first, detailed in \ref{sub:explicit}, using synthetically generated social networks to capture individuals' interactions; the second, discussed in \ref{sub:implicit}, focusing on a population - whose social structure is not given - stratified starting from Italian census data.

  \begin{figure}
\centering
 \subfloat[\textbf{SEIR}]{\includegraphics[scale=0.18]{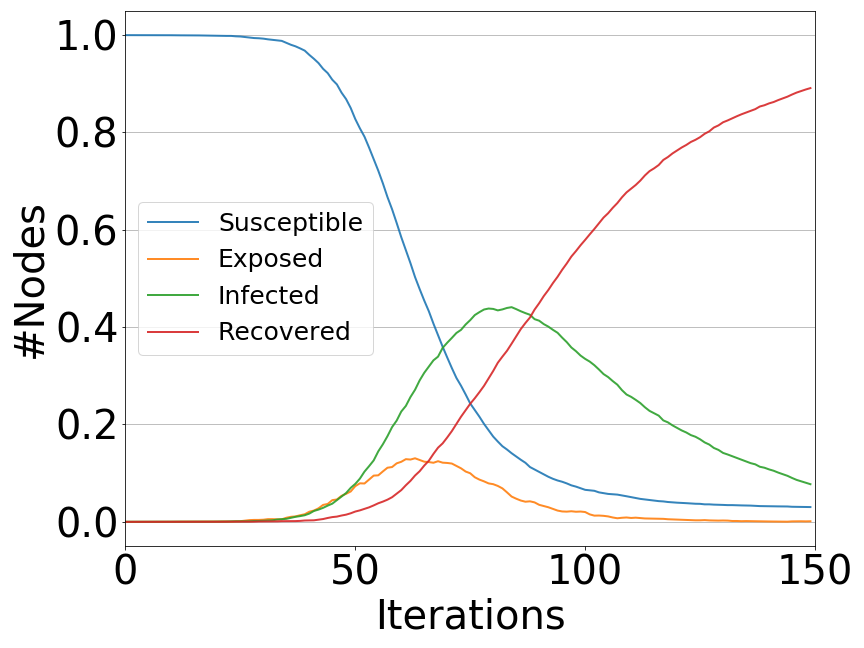}}
 \subfloat[\textbf{SEIS}]{\includegraphics[scale=0.18]{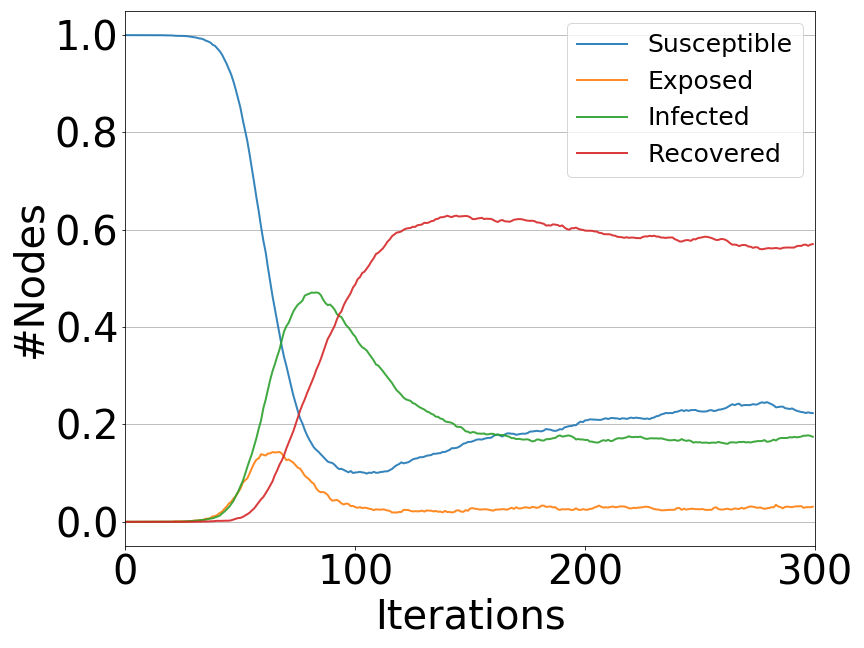}}
 \subfloat[\textbf{UTR}]{\includegraphics[scale=0.18]{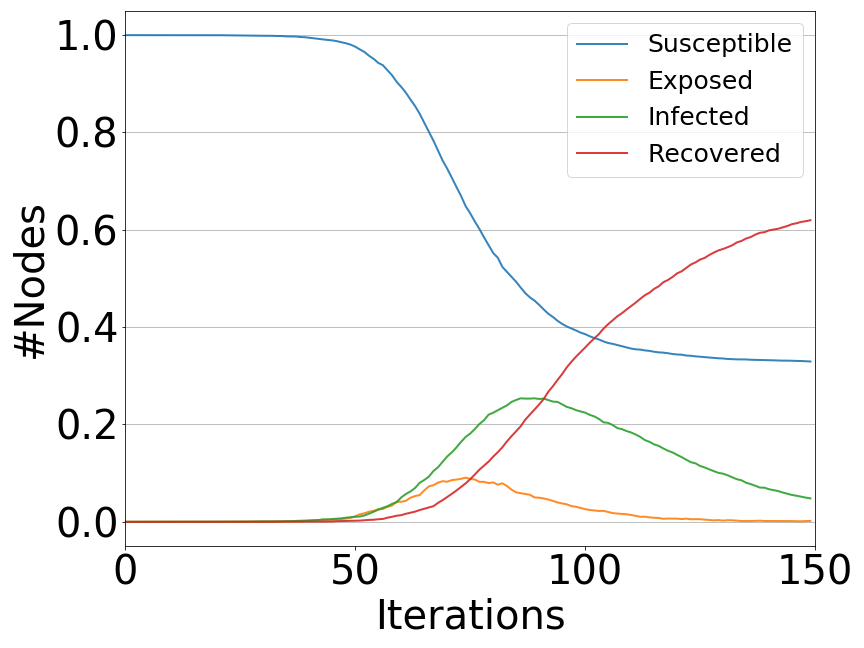}}
 \quad
 \subfloat[\textbf{UTLR}]{\includegraphics[scale=0.18]{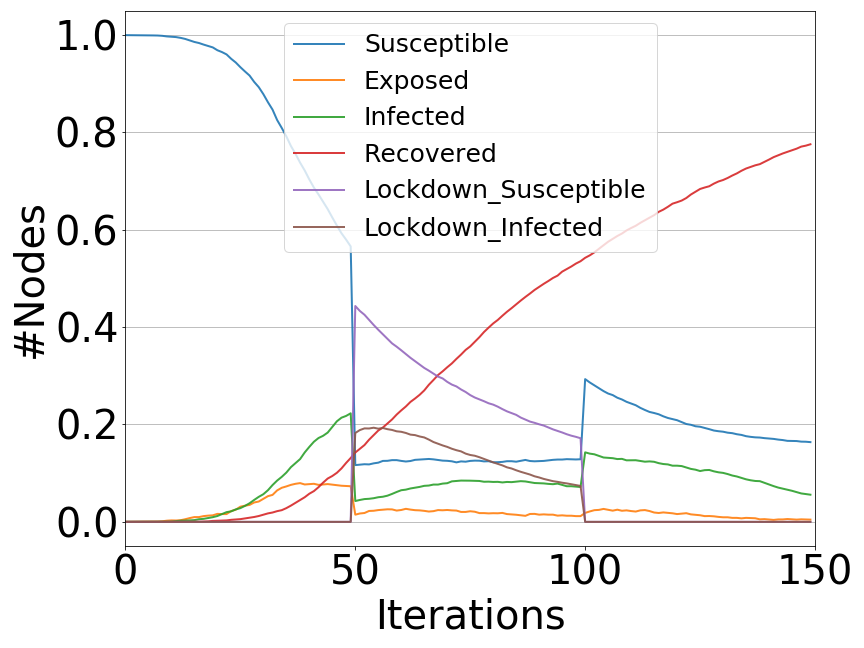}}
 \subfloat[\textbf{UTLDR}]{\includegraphics[scale=0.18]{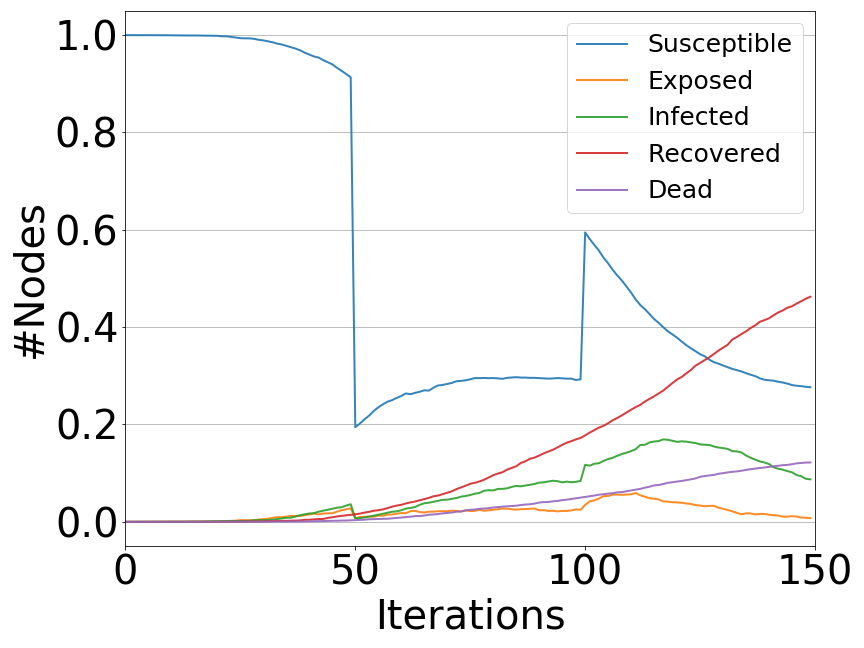}}
 \subfloat[\textbf{ICU}]{\includegraphics[scale=0.18]{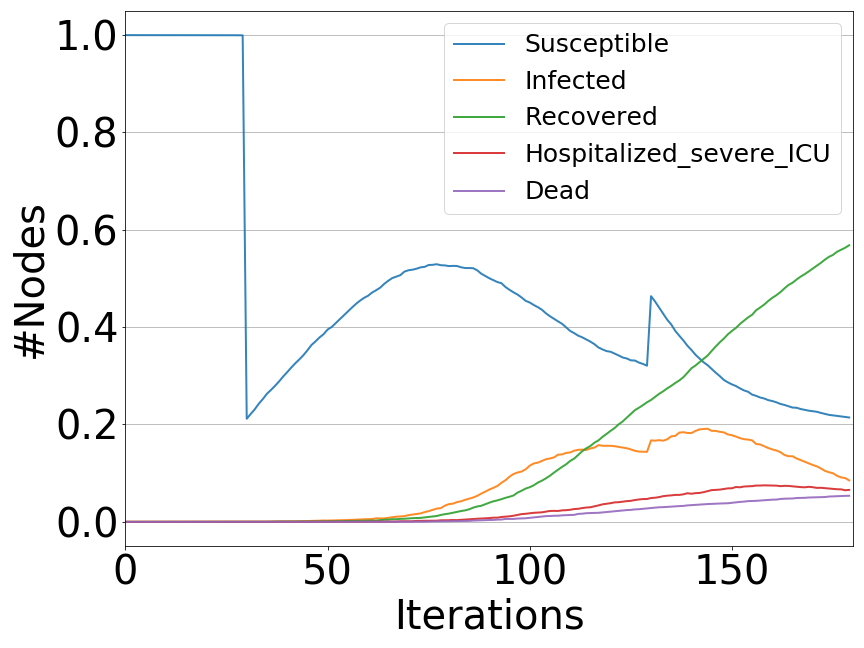}}
  \caption{Experiments on the BA model. (a) The simplest SEIR + (b) re-infection allowed transition; (c) Testing and (d) Lockdown scenarios + (e) dead-recovered distinction + (f) ICU availability.}
\label{ba_analysis}
  \end{figure}

\subsection{Explicit Network Structure}
\label{sub:explicit}

\begin{figure}
\centering
 \subfloat[\textbf{SEIR}]{\includegraphics[scale=0.18]{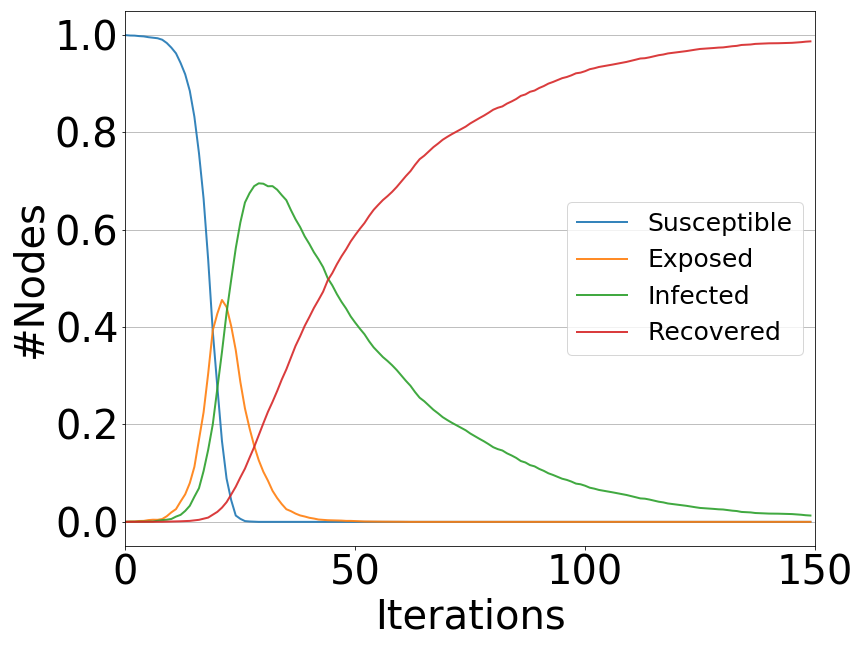}}
 \subfloat[\textbf{SEIS}]{\includegraphics[scale=0.18]{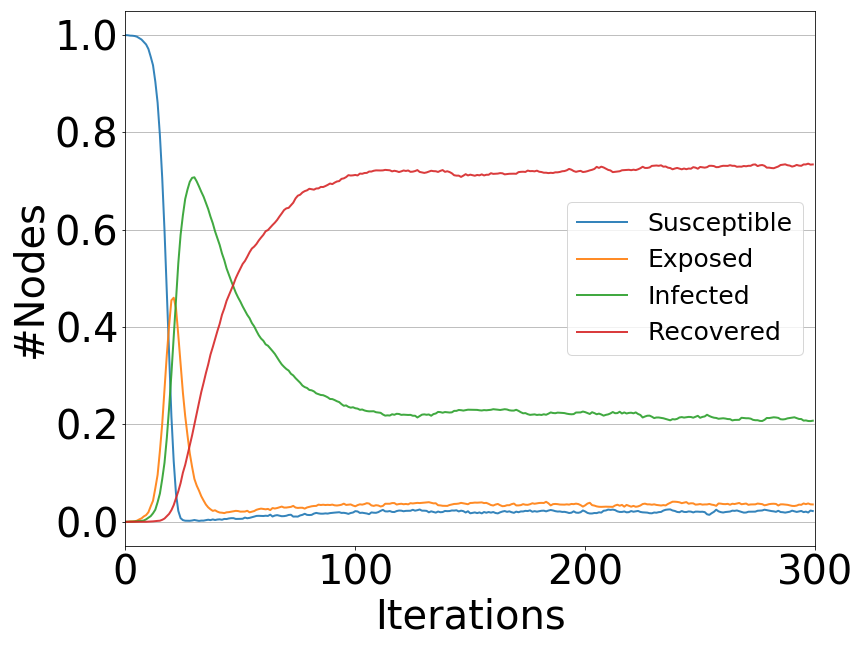}}
 \subfloat[\textbf{UTR}]{\includegraphics[scale=0.18]{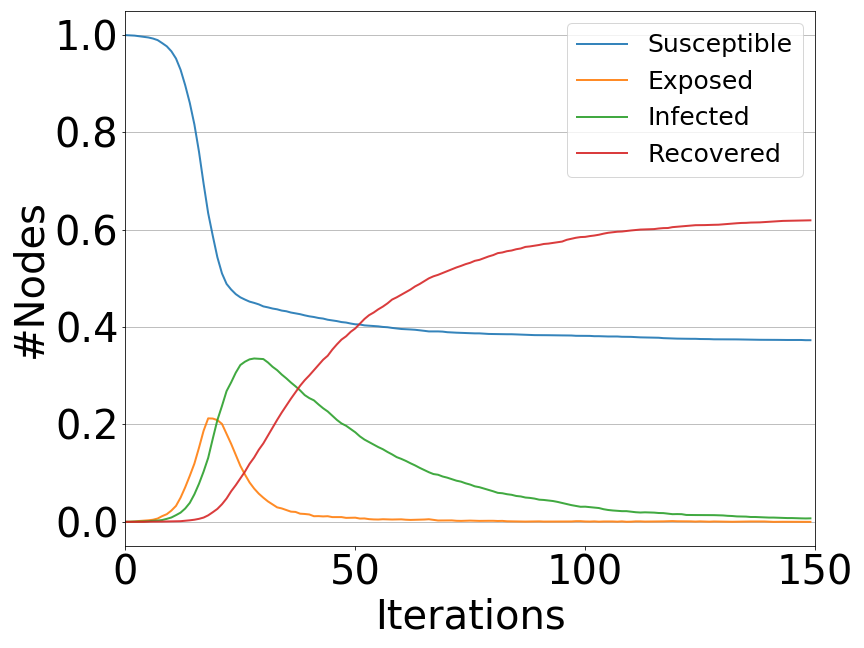}}
 \quad
 \subfloat[\textbf{UTLR}]{\includegraphics[scale=0.18]{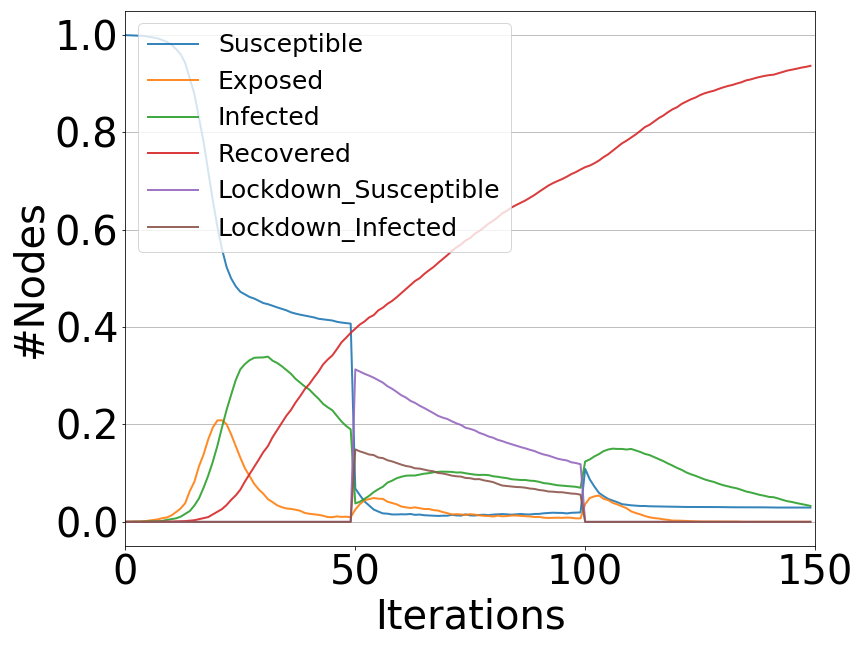}}
 \subfloat[\textbf{UTLDR}]{\includegraphics[scale=0.18]{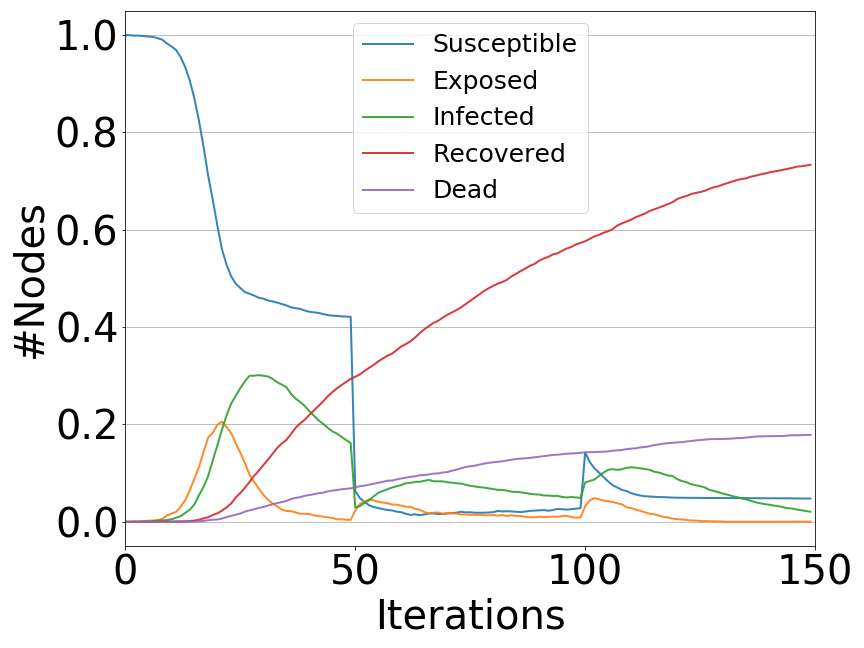}}
 \subfloat[\textbf{ICU}]{\includegraphics[scale=0.18]{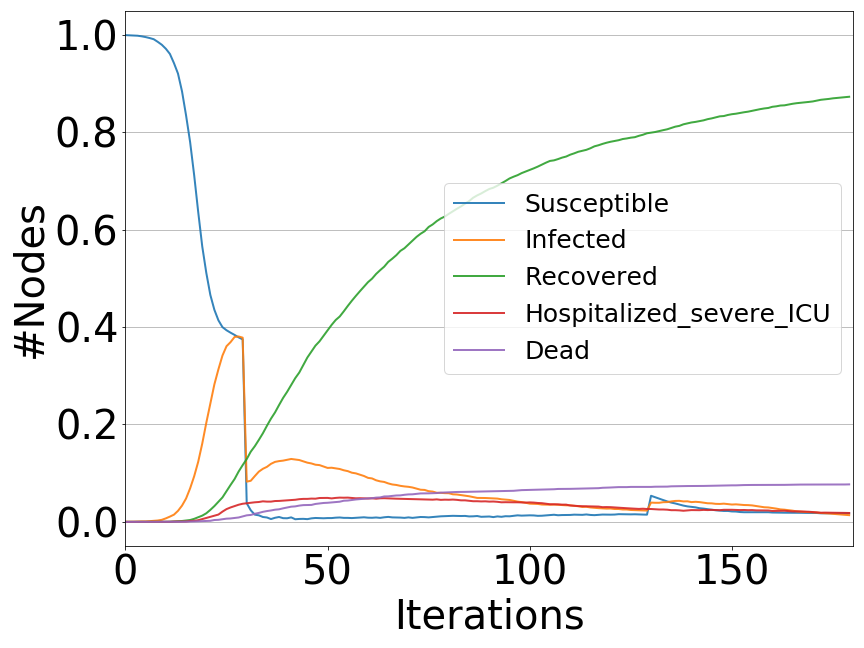}}
  \caption{Experiments on the ER model: (a) The simplest SEIR + (b) re-infection allowed transition; (c) Testing and (d) Lockdown scenarios + (e) dead-recovered distinction + (f) ICU availability.}
\label{er_analysis}
  \end{figure}

In Figure \ref{ba_analysis} and Figure \ref{er_analysis}, we show the diffusion trends obtained by simulating alternative models built on top of the proposed framework.
All simulations are executed assuming the underlying social structure as generated by the Barabási-Albert \cite{barabasi1999emergence} (henceforth, BA) and the Erdös-Rényi models \cite{erdds1959random} (henceforth, ER), each one composed by $N=5000$ nodes.
For the sake of simplicity, we do not integrate human mobility and population stratification in the reported case studies.

We set an initial fraction of infected nodes to $0.0001$, and simulate $150$ iterations - except for the SEIR model, where the number of iterations is extended to $300$, to observe better a possible re-infection effect. 
Moreover, while modeling the compartments where a lockdown is included (i.e., UTLR and the remaining incremental modules), the first $50$ iterations are run without any social distancing policy in place, the following $50$ imposing lockdown restrictions and, finally, a release of such policies during the remaining iterations.

The following sets of parameters are used for testing the compared models (models are reported in incremental order, the $nth$ one inheriting the parameter values of all the previous):
\begin{itemize}
    \item[SEIR:] $\beta=0.02$, $\sigma=0.2$, $\gamma=0.03$;
    \item[SEIS:] $s=0.01$ (partial immunization, allows $R\rightarrow S$);
    \item[UTR:] $\vartheta_E, \vartheta_I = 0.01$, $\kappa_E, \kappa_I = 0.05$, $\gamma_T = \gamma$;
    \item[UTLR:] $\tau=0.8$, $\mu=0.01$;
    \item[UTLDR:] $\omega, \omega_t=0.05$;
    \item[ICU:] $b=N$, $\iota=0.8$;
\end{itemize}

Please, note that the selected values are chosen for testing how the framework works and they do not reflect any real-world possible scenario.
In detail, while modeling the simplest SEIR model, we used a set of parameters that take an exposition period of $5$ days (i.e., $1/\sigma$) into account, and similar infection $\beta$ and recovery $\gamma$ rates. 
Among the two structures, the breakout is quicker in the ER model than in the BA one (Figures \ref{ba_analysis}(a) and \ref{er_analysis}(a)), and such a difference continues to be observed among the other incremental modules.
However, adding a re-infection parameter $s$, the effect is visible in the only BA model (Figures \ref{ba_analysis}(b) and \ref{er_analysis}(b)).

Introducing testing allows us to shift to the UTR module, where exposed and infected agents can be tested with probabilities $\vartheta_E$ and $\vartheta_I$, and with the possibility to get false positives with probability $\kappa_E$ and $\kappa_I$.
We used the same parameter values for our synthetic networks without differentiating between exposed and infected people, also considering a very low probability of getting false positives.
For simplicity, we set the recovery rate $\gamma_T$ equal to $\gamma$. 
In any case, the effect of quarantine (i.e., identified exposed and identified infected agents) is visible by observing the \textit{plateau} of the susceptible population curve (Figures \ref{ba_analysis}(c) and\ref{er_analysis}(c)), in both the two networks.

We then shift to the UTLR model compartment by specifying the two parameters that regulate social distancing/lockdown, i.e., the population adherence to the restrictions $\tau$ and the escape probability of the social distanced individuals $\mu$;
Here, we decide to report a scenario with high adherence to the imposed restrictions and a low escape probability to make more visible the differences between the first $50$ iterations and the remaining ones where lockdown restrictions are imposed.
Susceptible and infected social distanced agents permit to stop the breakout, then the infection can restart (Figures \ref{ba_analysis}(d) and \ref{er_analysis}(d)). 
No significant differences are observed among the two different topologies used in our experiments, except for the already underlined faster breakout in the ER model.

To finally introducing the UTLDR model, we specify the expected death rate.
We decide to set a particularly high death rate $\omega=0.05$ (imposing a same value for $\omega_T$) so to observe better a sharp death trend - as underlined in Figures \ref{ba_analysis}(e) and \ref{er_analysis}(e).
Finally, in Figures \ref{ba_analysis}(f) and \ref{er_analysis}(f), a simple parameter setting for ICU modeling is used, i.e., the number of ICU $b$ as the same as the agents in the networks, and a high percentage of ICU needs $\iota$.

\subsection{Implicit Network Structure}
\label{sub:implicit}

Differently from the previously discussed scenario, we assume that the social graph is not known in advance.
However, we also assume the presence of a carefully stratified set of agents designed to approximate the whole population of a given geographical area.
In particular, we perform our simulations on 3,73 million agents stratified to match an Italian region's population, Tuscany.
The population has been stratified by leveraging official census data as provided by ISTAT\footnote{http://dati.istat.it/}. 
In particular, the following dimensions have been used to characterize each agent and assign it to the proper geographic/social clusters:
\begin{itemize}
    \item Age, gender, household size distribution at the census cell level;
    \item Workplace (public/private sector and NACE code \cite{schnabl2013statistical}): number and size distribution at municipality level;
    \item Unemployment rate - stratified by age - at province level;
    \item Schools (by order): distribution of the number of classes and students (by age) at the municipality level.
\end{itemize}
Moreover, origin-destination matrices were simulated (due to lack of precise data) to consider mobility probabilities among a three-tiered hierarchy composed of census cells, municipalities, and provinces.
The simulated destination matrices rely only on geographical proximity, not on observed mobility fluxes.

The data used for this case study (along with stratified populations for all Italian regions), as well as the fine-tuned implementation of UTLDR, adapted to handle implicit network structure and human mobility, are available on a dedicated GitHub repository\footnote{https://github.com/KDDComplexNetworkAnalysis/UTLDR}.

\begin{figure}
    \centering
    \subfloat[]{\includegraphics[scale=0.16]{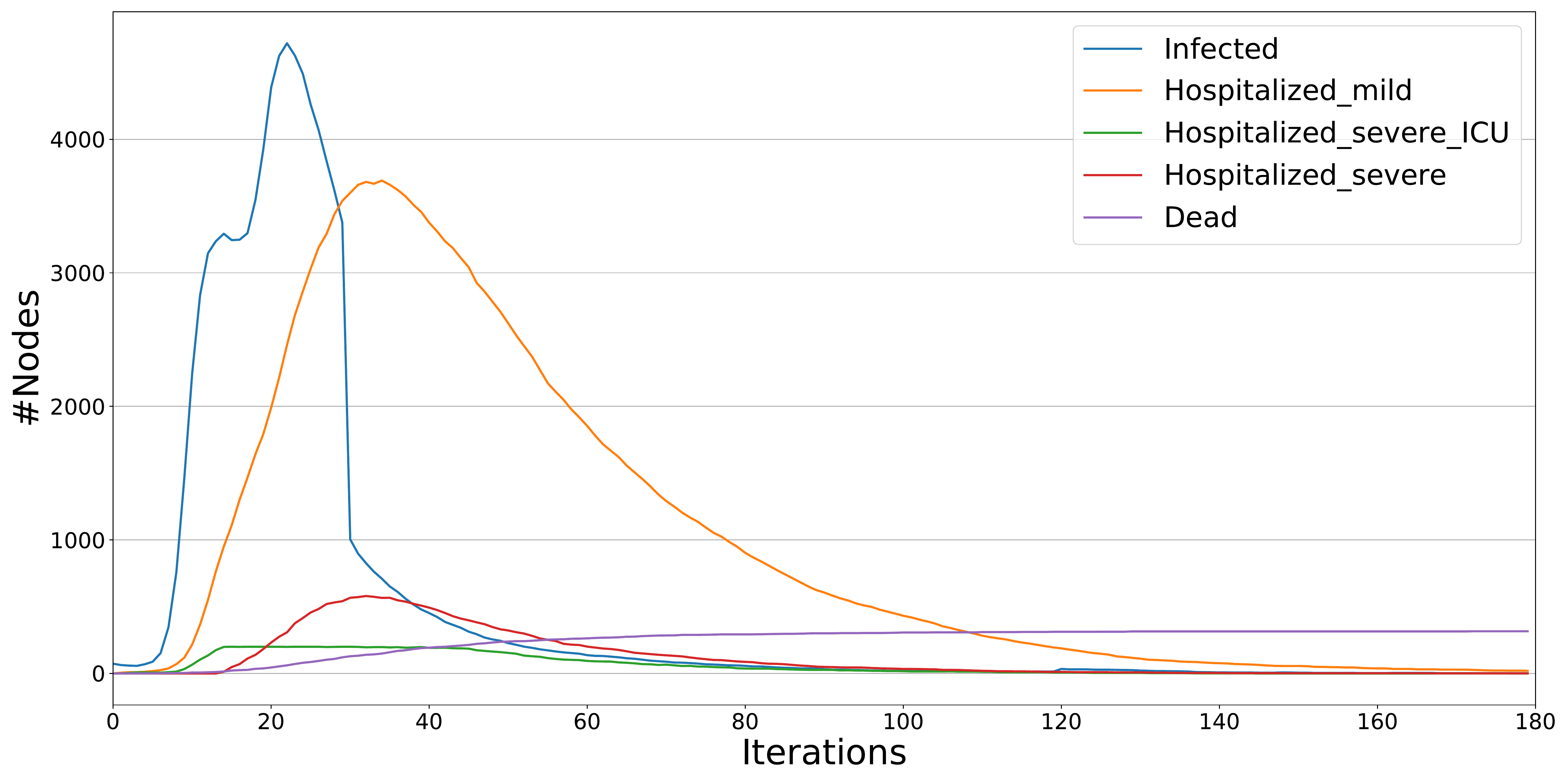}}
    \subfloat[]{\includegraphics[scale=0.16]{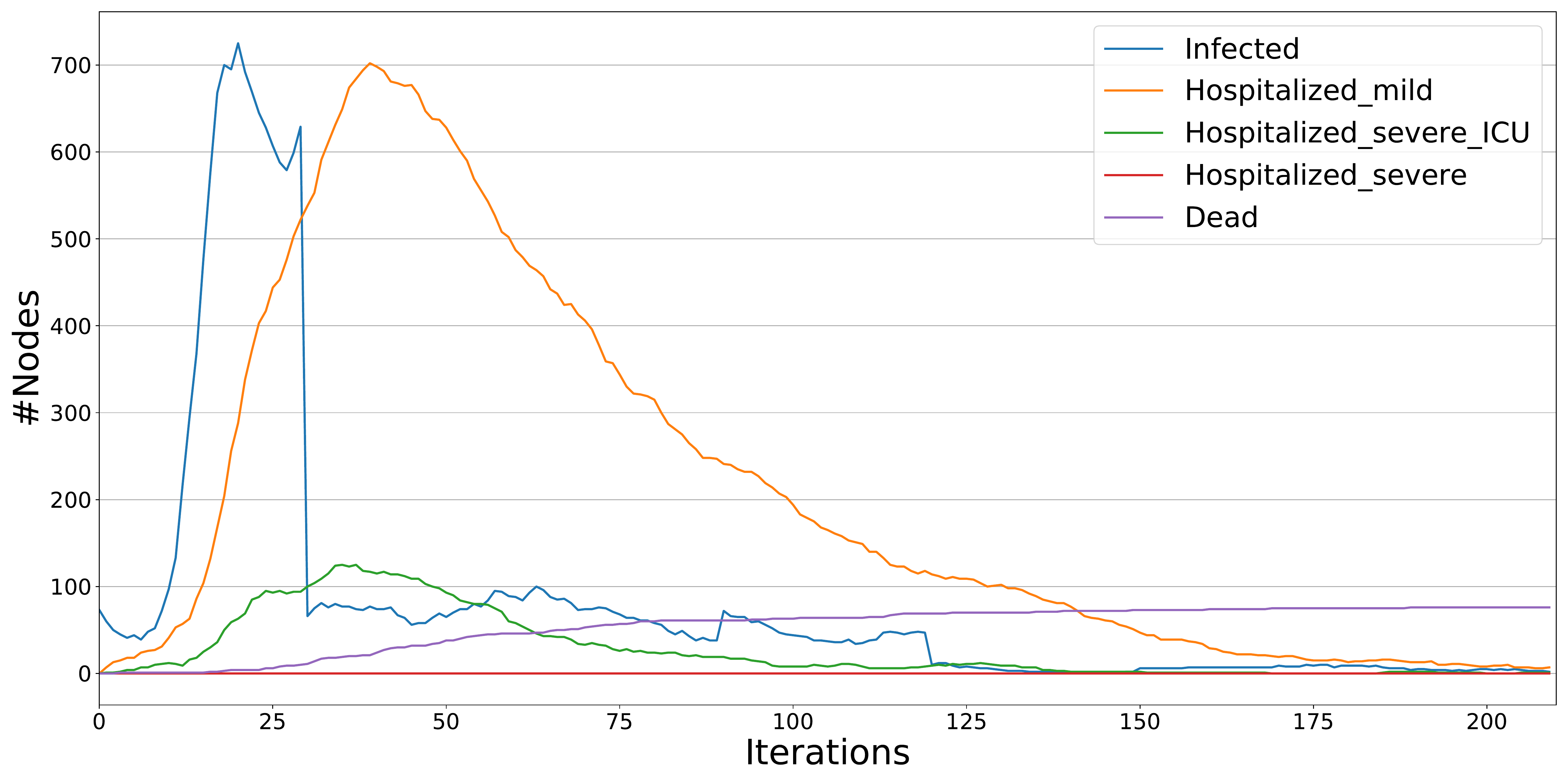}}\\
    \subfloat[]{\includegraphics[scale=0.16]{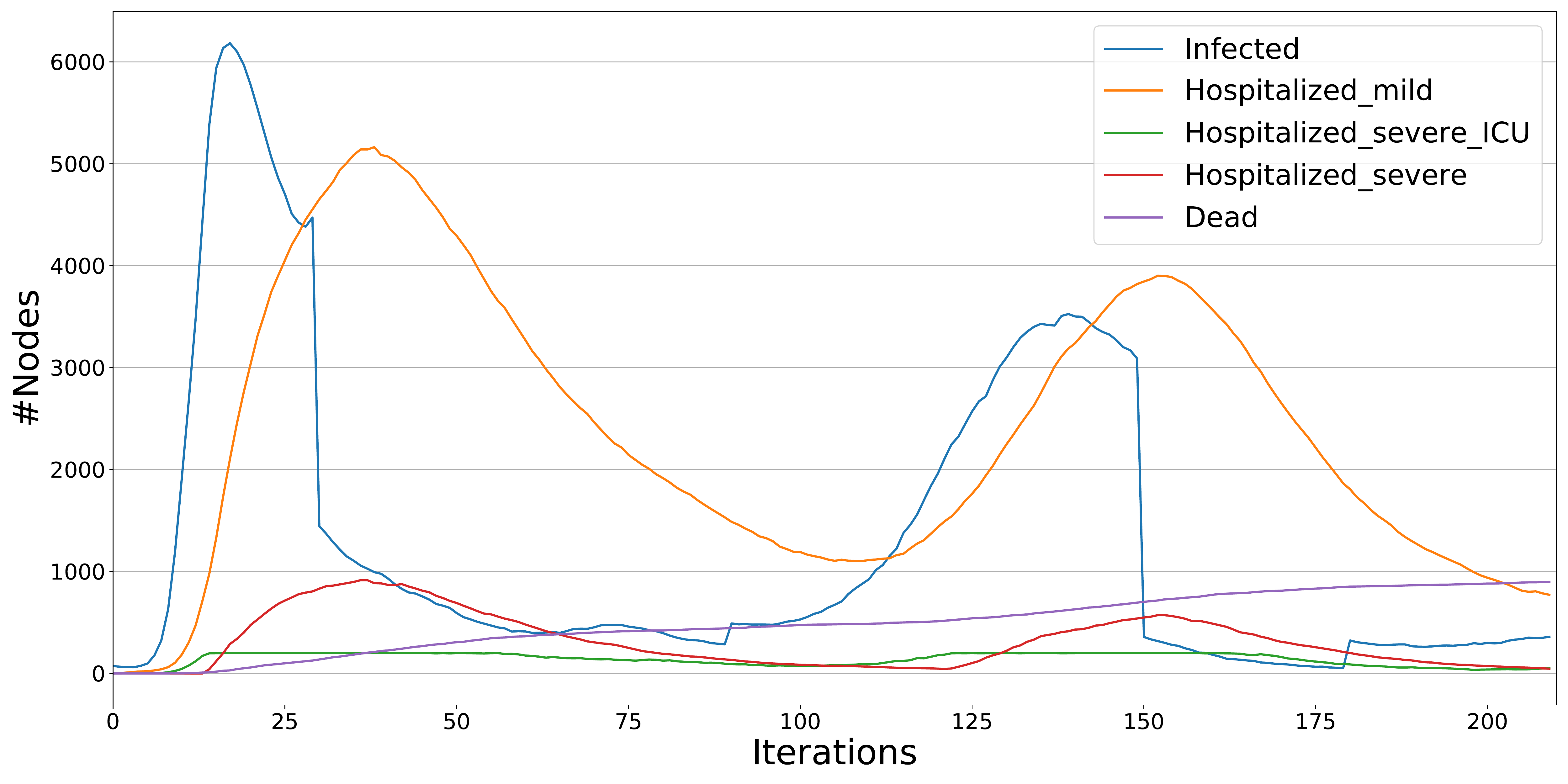}}
    \subfloat[]{\includegraphics[scale=0.16]{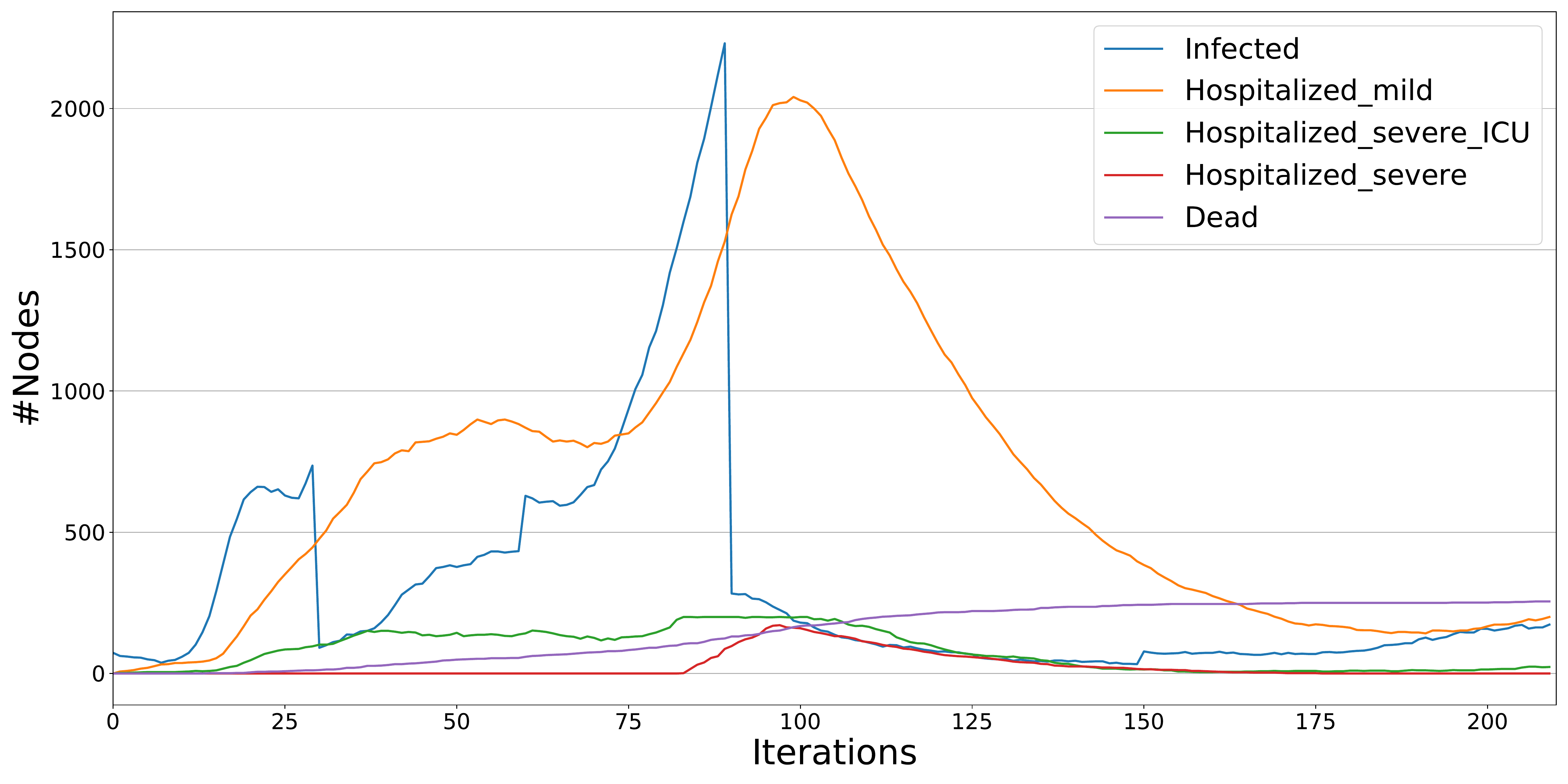}}
   
    \caption{Diffusion trends for the Tuscany case study. All scenarios start with a setup phase of 30 iterations, during which only UTDR compartments are active. (a) A single lockdown of 90 iterations is activated; (b) two consecutive lockdown of length 60 and 30 iterations respectively are activated - separated by 30 iterations of UTDR; (c) The same setting of the previous scenario but the separation among consecutive lockdown is set to 60 iterations; (d) Same setting of (b) but lockdown lengths are switched.}
    \label{tusc_analysis}
\end{figure}

In Figure \ref{tusc_analysis}(a-b), we report the diffusion trends for two different scenarios designed with UTLDR.
Both figures refer to the same model, the only significant variation lying on the temporal schedule of lockdowns.
In particular, the implemented model is completely specified by the following parameter settings:
\begin{itemize}
    \item Initial infected population: $0.00002\%$ ($\simeq 80$ individuals);
    \item SEIR parameters: $\beta=0.006$, $\sigma=0.25$, $\gamma=\gamma_T=0.04$;
    \item Testing parameters: $\vartheta_E=\kappa_e=0$ (no testing on exposed), $\vartheta_I=\kappa_i=0.1$;
    \item Tracing: $T_{tracing}=0$ (no tracing);
    \item Lethality: (real) $\omega=0.001$, (observed) $\omega_T=0.0015$;
    \item Long-range interactions: $p=0.008$;
    \item ICU: $\iota=0.2$, $b=200$ (pre-lockdown), $b=400$ (during)
    \item Lockdown: $\tau=0.9$.
\end{itemize}

To summarize, the designed model activates both hospitalization and lockdown compartments and allows long-range interactions. Moreover, during lookdowns, mobility is allowed only within the municipality boundaries, and the only categories not affected by lockdown policies are the Health workers.

The parameters of the activity driven model (the degree of activeness of each individual) are set by stratifying per age groups and social context (home census cell, workplace, school).
For instance, we assume for agents with age within [10-25] the following activeness scores ["workplace"=0, "home census cell"=0.05, "school"=0.9], while for those ones in the age range [25-50] the scores ["workplace"=0.4, "home census cell"=0.1, "school"=0].

Leveraging the described model, we design four different scenarios, each composed of 180 iterations (one per full day).
For the sake of simplicity, we report the trends only for a few compartments, namely: Infected (undetected), Hospitalized mild (quarantined), Hospitalized severe ICU, Hospitalized severe (patients requiring ICU but assigned to standard care for lack of beds), and Dead.
The total of ill individuals - during each iteration - is given by the sum of the first four compartments.
Exposed and Recovered populations are omitted so to increase readability.

In the first scenario, shown in Figure \ref{tusc_analysis}(a), after an initialization phase of 30 iterations - common to all proposed simulations -, where only testing is enabled, a lockdown of 90 iterations is imposed.
In the second scenario (Figure \ref{tusc_analysis}(b)), two consecutive lockdowns are activated: one right after the initialization phase - for 60 iterations -, the other 30 iterations after the deactivation of the previous one (and lasting for 30 iterations).
The third scenario follows the same pattern of the previous one while increasing the distance among the two imposed restrictions from 30 to 60 iterations (Figure \ref{tusc_analysis}(c)).
Finally, in the fourth scenario, the lockdown strategy designed in the second one are switched: imposing at first a 60 iterations lockdown and then, after 30 iterations, a new shorter one of 30 iterations (Figure \ref{tusc_analysis}(d)).

The reported trends clearly highlight how the length and scheduling of public interventions deeply affect the persistence of the simulated epidemic process - even while maintaining fixed the model describing it.
In the first and second scenarios, after a long closure, the epidemic completely dies out; conversely, in the third and fourth scenarios, the repeated lockdown is not enough to stop the diffusion. 
It is worth noticing that due to the stochasticity introduced by heterogeneous mixing patterns (and random infection seeds) the overall scale of the reported trends might vary from an execution to another: however, such variations in volume do not deeply affect the observed trends shape, producing only minor point-wise fluctuations.

\section{Discussion and Future Works}
\label{sec:future}
This paper introduced a framework, namely UTLDR, to allow an incremental description of compartmental epidemic models incorporating diffusion mitigation strategies. 
UTLDR segments the compartments it provides in five macro classes (Undetected, Tested, Lockdown, Recovered, and Dead), each one identifying a set of interventions/outcomes.
After discussing the compartments, transition rules among them, and controlling parameters, we provided a few examples of models that UTLDR allows to build - ranging from simple Testing and Hospitalization to Lockdown and Social Distancing.
Moreover, we also showed how additional compartments could be easily added to UTLDR models to simulate vaccination strategies and inefficient corpse disposal.

We formulated UTLDR in a conventional mean-field scenario (as reported in Appendix \ref{appendix}): however, to account for heterogeneous mixing-patterns, we also discussed its extension to complex network topologies.
We modeled such a scenario with an activity-driven network approach, allowing as inputs both explicit and implicit social tissues that dynamically update during the model simulation.
Moreover, we designed UTLDR to allow stratified parameter settings on top of population characteristics (e.g., age, gender\dots) and, at the same time, to incorporate aggregate human mobility information (as origin-destination matrices) to better account for geographic contacts limitations.

Finally, we provided case studies to qualitatively discuss a subset of the models built on top of UTLDR, focusing on the framework explicit and implicit network extension.

As future work, we plan to extend the Tuscany case study analysis to other Italian regions and define a model focused on SARS-CoV-2 scenario generation.
Moreover, we plan to release an online dashboard to support a visual setup of UTLDR models simulation and visual inspection of their results.

\subsubsection*{Acknowledgement}
This work is supported by the scheme 'INFRAIA-01-2018-2019: Research and Innovation action', Grant Agreement n. 871042 'SoBigData++: European Integrated Infrastructure for Social Mining and Big Data Analytics'

\subsubsection*{Author Contributions}
G.R. designed and coded the framework, L.M. formulated the mean-field description, S.C. and V.M. performed the experiments. All authors contributed to paper writing.

\bibliographystyle{unsrt}  
\bibliography{references}  

\newpage
\appendix
\section{Mean Field Formulation}
\label{appendix}
\subsection{Testing and Quarantine}
 In the closed population, with no births or deaths, the UTR model can be described as follows:
    \begin{align*}
       & \frac{dS}{dt}= \frac{-\beta SI}{N} \\
    &    \frac{dE}{dt}= \frac{\beta SI}{N} - \sigma E- \vartheta_E \kappa_E E \\
    & \frac{dI}{dt}=  \sigma E- \vartheta_I \kappa_I I - \gamma I \\
    &    \frac{dE_T}{dt}= \vartheta_E \kappa_E E - \sigma E_T 
    \\
        &    \frac{dI_T}{dt}=  \sigma E_T + \vartheta_I \kappa_I I - \gamma_T I_T\\
        & \frac{dR}{dt} = \gamma I + \gamma_T I_T
    \end{align*}
where $N= S+E+E_T+I+I_T+R$ is the total population.

\subsection{Lockdown}
   In the closed population, with no births or deaths, the UTLR model can be described as follows:
    \begin{align*}
        & \frac{dS}{dt}= \frac{-\beta SI}{N} -\tau S + \mu S_L \\
          & \frac{dS_L}{dt}= \frac{-\beta S_L I_L}{N} +\tau S - \mu S_L \\
    &    \frac{dE}{dt}= \frac{\beta SI}{N} - \sigma E- \vartheta_E \kappa_E E \\
      &    \frac{dE_L}{dt}= \frac{\beta S_L I_L}{N} - \sigma E_L- \vartheta_E \kappa_E E_L \\
          &    \frac{dE_T}{dt}= \vartheta_E \kappa_E E - \sigma E_T 
          +  \vartheta_E \kappa_E E_L\\
    & \frac{dI}{dt}=  \sigma E- \vartheta_I \kappa_I I - \gamma I\\
    & \frac{dI_L}{dt}=  \sigma E_L- \vartheta_I \kappa_I I_L - \gamma I_L \\
        &
        \frac{dI_T}{dt}=  \sigma E_T + \vartheta_I \kappa_I I - \gamma_T I_T +\vartheta_I \kappa_I I_L\\
        & \frac{dR}{dt} = \gamma (I+I_L) + \gamma_T I_T
    \end{align*}
where $N= S+S_L+E+E_L+E_T+I+I_L+I_T+R$ is the total population.

\subsection{Dead state}

The UTLDR model considering the Dead compartment becomes:
    \begin{align*}
        & \frac{dS}{dt}= \frac{-\beta SI}{N} -\tau S + \mu S_L \\
          & \frac{dS_L}{dt}= \frac{-\beta S_L I_L}{N} +\tau S - \mu S_L \\
    &    \frac{dE}{dt}= \frac{\beta SI}{N} - \sigma E- \vartheta_E \kappa_E E \\
      &    \frac{dE_L}{dt}= \frac{\beta S_L I_L}{N} - \sigma E_L- \vartheta_E \kappa_E E_L \\
          &    \frac{dE_T}{dt}= \vartheta_E \kappa_E E - \sigma E_T
          +  \vartheta_E \kappa_E E_L\\
    & \frac{dI}{dt}=  \sigma E- \vartheta_I \kappa_I I - \gamma I- \omega I\\
    & \frac{dI_L}{dt}=  \sigma E_L- \vartheta_I \kappa_I I_L - \gamma I_L - \omega I_L\\
        &    \frac{dI_T}{dt}=  \sigma E_T + \vartheta_I \kappa_I I - \gamma_T I_T +\vartheta_I \kappa_I I_L - \omega_T I_T\\
        & \frac{dR}{dt} = \gamma (I+I_L) + \gamma_T I_T \\
        & \frac{dD}{dt} =\omega (I + I_L)+ \omega_T I_T
    \end{align*}
where $N= S+S_L+E+E_L+E_T+I+I_L+I_T+R+D$ is the total population.

\subsection{ICU limitations}

The UTLDR model integrating ICU limitations becomes:
    \begin{align*}
        & \frac{dS}{dt}= \frac{-\beta SI}{N} -\tau S + \mu S_L \\
          & \frac{dS_L}{dt}= \frac{-\beta S_L I_L}{N} +\tau S - \mu S_L \\
    &    \frac{dE}{dt}= \frac{\beta SI}{N} - \sigma E- \vartheta_E \kappa_E E \\
      &    \frac{dE_L}{dt}= \frac{\beta S_L I_L}{N} - \sigma E_L- \vartheta_E \kappa_E E_L \\
          &    \frac{dE_T}{dt}= \vartheta_E \kappa_E E - 2\sigma E_T +  \vartheta_E \kappa_E E_L\\
    & \frac{dI}{dt}=  \sigma E- \vartheta_I \kappa_I I - \gamma I- \omega I\\
    & \frac{dI_L}{dt}=  \sigma E_L- \vartheta_I \kappa_I I_L - \gamma I_L - \omega I_L\\
        &    \frac{dI_T}{dt}=  \sigma E_T + \vartheta_I \kappa_I I -ibI_T- \gamma I_T  - \omega I_T\\
            &    \frac{dH_T}{dt}=  \sigma E_T + \vartheta_I \kappa_I I_L +ibI_T- \gamma_T H_T  - \omega_T H_T\\
        & \frac{dR}{dt} = \gamma (I+I_L+I_T) + \gamma_T H_T \\
        & \frac{dD}{dt} =\omega (I+I_L+I_T) + \omega_T H_T
    \end{align*}
where $N= S+S_L+E+E_L+E_T+I+I_L+I_T+H_T+R+D$ is the total population.

\subsection{Corpse Disposal Efficiency}

The UTLDR model integrating corpse disposal becomes:
    \begin{align*}
        & \frac{dS}{dt}= \frac{-\beta SI}{N} -\tau S + \mu S_L +z D \\
          & \frac{dS_L}{dt}= \frac{-\beta S_L I_L}{N} +\tau S - \mu S_L \\
    &    \frac{dE}{dt}= \frac{\beta SI}{N} - \sigma E- \vartheta_E \kappa_E E \\
      &    \frac{dE_L}{dt}= \frac{\beta S_L I_L}{N} - \sigma E_L- \vartheta_E \kappa_E E_L \\
          &    \frac{dE_T}{dt}= \vartheta_E \kappa_E E - 2\sigma E_T +  \vartheta_E \kappa_E E_L\\
    & \frac{dI}{dt}=  \sigma E- \vartheta_I \kappa_I I - \gamma I- \omega I\\
    & \frac{dI_L}{dt}=  \sigma E_L- \vartheta_I \kappa_I I_L - \gamma I_L - \omega I_L\\
        &    \frac{dI_T}{dt}=  \sigma E_T + \vartheta_I \kappa_I I -ibI_T- \gamma I_T  - \omega I_T\\
            &    \frac{dH_T}{dt}=  \sigma E_T + \vartheta_I \kappa_I I_L +ibI_T- \gamma_T H_T  - \omega_T H_T\\
        & \frac{dR}{dt} = \gamma (I+I_L+I_T) + \gamma_T H_T \\
        & \frac{dD}{dt} =\omega (I+I_L+I_T) + \omega_T H_T -z D
    \end{align*}
where $N= S+S_L+E+E_L+E_T+I+I_L+I_T+H_T+R+D$ is the total population. 

\subsection{Partial Immunity}

The UTLDR model integrating partial immunity becomes:
    \begin{align*}
        & \frac{dS}{dt}= \frac{-\beta SI}{N} -\tau S + \mu S_L +z D +s R \\
          & \frac{dS_L}{dt}= \frac{-\beta S_L I_L}{N} +\tau S - \mu S_L \\
    &    \frac{dE}{dt}= \frac{\beta SI}{N} - \sigma E- \vartheta_E \kappa_E E \\
      &    \frac{dE_L}{dt}= \frac{\beta S_L I_L}{N} - \sigma E_L- \vartheta_E \kappa_E E_L \\
          &    \frac{dE_T}{dt}= \vartheta_E \kappa_E E - 2\sigma E_T +  \vartheta_E \kappa_E E_L\\
    & \frac{dI}{dt}=  \sigma E- \vartheta_I \kappa_I I - \gamma I- \omega I\\
    & \frac{dI_L}{dt}=  \sigma E_L- \vartheta_I \kappa_I I_L - \gamma I_L - \omega I_L\\
        &    \frac{dI_T}{dt}=  \sigma E_T + \vartheta_I \kappa_I I -ibI_T- \gamma I_T  - \omega I_T\\
            &    \frac{dH_T}{dt}=  \sigma E_T + \vartheta_I \kappa_I I_L +ibI_T- \gamma_T H_T  - \omega_T H_T\\
        & \frac{dR}{dt} = \gamma (I+I_L+I_T) + \gamma_T H_T -s R\\
        & \frac{dD}{dt} =\omega (I+I_L+I_T) + \omega_T H_T -z D
    \end{align*}
where $N= S+S_L+E+E_L+E_T+I+I_L+I_T+H_T+R+D$ is the total population. 

\subsection{Vaccination}

The UTLDR model integrating vaccination strategies becomes:
    \begin{align*}
        & \frac{dS}{dt}= \frac{-\beta SI}{N} -\tau S + \mu S_L +z D +s R -v S + f V  \\
        & \frac{dV}{dt}= v S + v S_L -2 f V  \\
          & \frac{dS_L}{dt}= \frac{-\beta S_L I_L}{N} +\tau S - \mu S_L -v S_L + f V \\
    &    \frac{dE}{dt}= \frac{\beta SI}{N} - \sigma E- \vartheta_E \kappa_E E \\
      &    \frac{dE_L}{dt}= \frac{\beta S_L I_L}{N} - \sigma E_L- \vartheta_E \kappa_E E_L \\
          &    \frac{dE_T}{dt}= \vartheta_E \kappa_E E - 2\sigma E_T +  \vartheta_E \kappa_E E_L\\
    & \frac{dI}{dt}=  \sigma E- \vartheta_I \kappa_I I - \gamma I- \omega I\\
    & \frac{dI_L}{dt}=  \sigma E_L- \vartheta_I \kappa_I I_L - \gamma I_L - \omega I_L\\
        &    \frac{dI_T}{dt}=  \sigma E_T + \vartheta_I \kappa_I I -ibI_T- \gamma I_T  - \omega I_T\\
            &    \frac{dH_T}{dt}=  \sigma E_T + \vartheta_I \kappa_I I_L +ibI_T- \gamma_T H_T  - \omega_T H_T\\
        & \frac{dR}{dt} = \gamma (I+I_L+I_T) + \gamma_T H_T -s R\\
        & \frac{dD}{dt} =\omega (I+I_L+I_T) + \omega_T H_T -z D
    \end{align*}
where $N= S+S_L+V+E+E_L+E_T+I+I_L+I_T+H_T+R+D$ is the total population.

\end{document}